\documentclass{phb-proc4-auth}

\usepackage{graphicx}
\usepackage{amssymb}

\begin{document}
\begin{frontmatter}




\title{Design of the electronic structure of poly-MTO}

%
%
%
%
%
%

\author{R. Miller }
\author{Ch. Helbig }
\author{G. Eickerling}
\author{R. Herrmann}
\author{E. - W. Scheidt }
\author{W. Scherer }

%

\address{Chemische Physik und Materialwissenschaften,
Universit\"{a}t Augsburg, 86159 Augsburg, Germany}

%
%
%
%

\thanks{This work was supported by the SFB~484 of the Deutsche
Forschungsgemeinschaft (DFG).}

%
%
%
%



\begin{abstract}

Polymeric methyltrioxorhenium (poly-MTO) is the first member of a
new class of organometallic hybrids which adopts the structural
motives and physical properties of classical perowskites in two
dimensions. In this study we demonstrate how the electronic
structure of poly-MTO can be tailored by intercalation of organic
donor molecules such as tetrathiafulvalene (TTF). With increasing
donor intercalation the metallic behavior of the parent compound,
(CH$_{3}$)$_{0.92}$ReO$_{3} \cdot x\%$  TTF ($x = 0$) becomes
suppressed leading to an insulator at donor concentrations $x$
larger than 50. Specific heat, electric resistance and magnetic
susceptibility studies indicate that an increasing amount of TTF
causes the itinerant electrons of the poly-MTO matrix to localize.

\end{abstract}

%
%



\end{frontmatter}

%
%
%
%
%

Polymeric methyltrioxorhenium (poly-MTO),
(CH$_{3}$)$_{0.92}$ReO$_{3}$, represents the first example of a
conductive organometallic polymeric oxide. From analytical studies
it is known that $\approx$\,10\% of the methyl groups are missing
in comparison with the monomer. The Re d$^1$ centers which are
formed during demethylation are supposed to provide the itinerant
electrons of the conduction band, leading to metallic behavior
\cite{Fischer_1994,Hermann_1995}. Since its discovery by Herrmann,
Fischer and Scherer in 1992 \cite{Hermann_1992} several attempts
have been undertaken to use its unique chemical properties to
design new materials by manipulating its chemical composition.
Here we present for the first time successful approaches to
manipulate the electronic and magnetic structure of poly-MTO by
controlled intercalation employing the organic donor species
tetrathiafulvalene (TTF) attempting to increase the electronic
conductivity.

For the synthesis of poly-MTO, two methods have been established.
One consists in dissolving MTO (CH$_{3}$ReO$_{3}$) in water under
stirring at 80\,$^\circ$C for two days. However, this is not
applicable for the intercalation of TTF. Another method has been
established which is based on auto-polymerization of MTO in the
flux. In the same way finely ground mixtures of MTO are used to
intercalate TTF in a sealed ampoule at 120\,$^\circ$C during two
days. The intercalated samples (poly-MTO + $x\%$\,TTF) prepared by
this way form bronze- colored solids for low TTF concentrations
($x < 40$) and almost black powders for higher intercalation
ratios.

\begin{figure}[h]
\centering
\includegraphics[width=5.5 cm,clip]{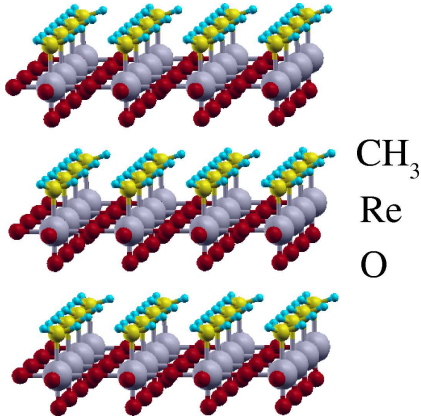}
\caption{Structural model of poly-MTO. Powder X-ray diffraction
studies suggest a two-dimensional \{ReO$_{2}$\}$_{\infty}$ layered
structure: Missing reflections related to interlayer stacking and
the asymmetric shape of the hk0 reflections indicate ordering
solely in two dimensions. Intercalation of TTF further reduces the
crystallinity of the samples.}
\label{fig1}                          
\end{figure}

X-ray powder diffraction measurements clearly show that
two-dimensional ordering of the \{ReO$_{2}$\}$_{\infty}$ framework
is preserved in all samples. Hence poly-MTO (Fig.~\ref{fig1})
displays a layered structure without significant interlayer
ordering. Intercalated TTF is assumed to settle in between the
layers to cause a decrease in crystallinity, which is indicated by
the powder diffraction studies.

\begin{figure}
\centering
\includegraphics[width=6.2 cm,clip]{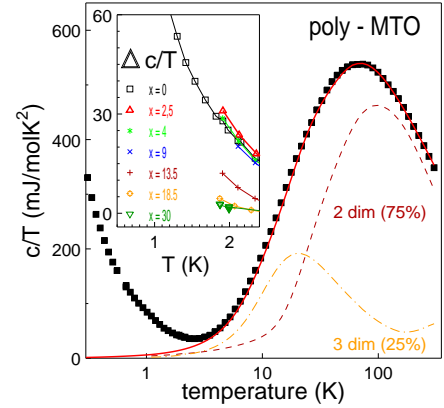}
\caption{Specific heat divided by temperature vs. log$T$.  The
phonon contribution (solid line) can be well parameterized using
75\% two - and 25\% three dimensional terms with
$\Theta_\mathrm{D}(2\mathrm{dim}) = 206$\,K and
$\Theta_\mathrm{D}(3\mathrm{dim}) = 66$\,K for the low temperature
region. The strong increase of the electronic part $\Delta c/T$
below 3\,K is plotted separately in the insert for selected TTF
samples. Here $x$ is the TTF to Re ratio in $\%$.}
\label{fig2}                          
\end{figure}

\begin{figure}
\centering
\includegraphics[width=6.5 cm,clip]{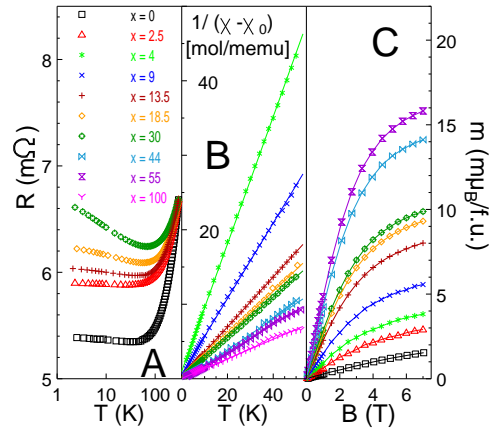}
\caption{The temperature dependence of the (A) resistance (note the
logarithmic scale), (B) the inverse magnetic susceptibility and (C)
the magnetic field dependence of the magnetization. All lines are
theoretical fits through the data and $x$ is the TTF to Re ratio in
$\%$. $\chi_{0}$ does not contain the diamagnetic contribution of
the core electrons which is estimated within an error of 10\%.}
\label{fig3}                          
\end{figure}

This two-dimensional character is also seen in the temperature
dependent phonon contribution calculated from specific heat
measurements (Fig.~\ref{fig2}). For the pure poly-MTO sample the
specific heat increases considerably below 3\,K giving rise to a
nuclear electric Schottky effect, but also disorder cannot be
ruled out at present. This significant upturn seems to be
suppressed with increasing TTF intercalation above $T = 2$\,K
(insert Fig.~\ref{fig2}). An estimated $\gamma \approx$
7\,$\pm$\,2\,mJ/molK$^2$ is in agreement with typical values
displayed by d band metals. Measurements of the resistance
(Fig.~\ref{fig3}A) display a decrease of the metallic behavior
with increasing amount of TTF. Furthermore the temperature
dependency of the resistance becomes reduced. This may be caused
by increasing amount of defects induced in the conducting layers,
in accord with our X-ray diffraction studies. Below 20 K an
logarithmic temperature dependency as it is known from
Kondo-impurity systems is observed. These results are opposite to
the common assumption that TTF as a donor acts as a source of
itinerant electrons.


To explain this unexpected behavior of the intercalated samples,
magnetic susceptibility studies have been performed
(Fig.~\ref{fig3}B). Below 50\,K $\chi$(T) is well accounted for by a
simple modified Curie-Weiss type behavior, $\chi (T) = C/(T-\Theta)
+ \chi_0$, yielding  $\chi_{0} \le $0.1 memu/mol and $\Theta \cong
0$~K indicating   no correlations of the residual localized d$^1$
electrons at Re positions. The effective paramagnetic moment
$\mu_\mathrm{eff}$ obtained from the Curie constant $C$ corresponds
to  0.05\% Re atoms carrying d$^1$ moments in the case of pure
poly-MTO and 2\% for samples with Re to TTF ratios of one.
Calculations from magnetization measurements (Fig.~\ref{fig3}C) lead
to the same amount of localized d$^1$ centers. On the other hand an
observed decrease of $\chi_{0}$ with TTF content exhibits a
reduction of the itinerant electron density which corroborates
nicely the conductivity data. In addition the Wilson-ratio $R =
\pi^2 k_\mathrm{B}^2 \chi_{0}/{2 \mu_0 \mu_\mathrm{B}^2 \gamma}
\cong 1$ as it is expected for non-magnetic metals.

Summarizing, an organic compound has been intercalated in a
polymeric organometallic oxide for the first time. TTF
intercalation in poly-MTO leads to a crossover from metallic to
insulating behavior.

%
%
%
%

%
%
%
%


\end{document}